\documentclass[preprint,showpacs,preprintnumbers,superscriptaddress,amsmath,amssymb]{revtex4}
\usepackage{graphicx}
\usepackage{dcolumn}
\usepackage{bm}
\usepackage{color}

\begin{document}

\title{Structural phase transition and superlattice misfit strain
of RFeAsO (R = La, Pr, Nd and Sm)}
\author{Alessandro Ricci}
\affiliation{Dipartimento di Fisica, Universit\`{a} di Roma ``La 
Sapienza", P. le Aldo Moro 2, 00185 Roma, Italy}
\author{Nicola Poccia}
\affiliation{Dipartimento di Fisica, Universit\`{a} di Roma ``La 
Sapienza", P. le Aldo Moro 2, 00185 Roma, Italy}
\author{B. Joseph}
\affiliation{Dipartimento di Fisica, Universit\`{a} di Roma ``La 
Sapienza", P. le Aldo Moro 2, 00185 Roma, Italy}
\author{Luisa Barba}
\affiliation{Institute of Crystallography, National Council of Research, 
Elettra, 34012 Trieste, Italy}
\author{G. Arrighetti}
\affiliation{Institute of Crystallography, National Council of Research, 
Elettra, 34012 Trieste, Italy}
\author{G. Ciasca}
\affiliation{Dipartimento di Fisica, Universit\`{a} di Roma ``La 
Sapienza", P. le Aldo Moro 2, 00185 Roma, Italy}
\author{J.-Q. Yan}
\affiliation{Divison of Materials Science and Engineering, Ames Laboratory, 
US-DOE, Ames, Iowa, 50011, USA}
\author{R. W. McCallum}
\affiliation{Divison of Materials Science and Engineering, Ames Laboratory, 
US-DOE, Ames, Iowa, 50011, USA}
\author{T. A. Lograsso}
\affiliation{Divison of Materials Science and Engineering, Ames Laboratory, 
US-DOE, Ames, Iowa, 50011, USA}
\author{N. D. Zhigadlo}
\affiliation{Laboratory of Solid State Physics, ETH Zurich, 8093 Zurich, 
Switzerland}
\author{J. Karpinski}
\affiliation{Laboratory of Solid State Physics, ETH Zurich, 8093 Zurich, 
Switzerland}
\author{Antonio Bianconi}
\email[Corresponding author: ]{Antonio.Bianconi@roma1.infn.it}
\affiliation{Dipartimento di Fisica, Universit\`{a} di Roma ``La 
Sapienza", P. le Aldo Moro 2, 00185 Roma, Italy}

\begin{abstract}
The tetragonal-to-orthorhombic structural phase transition (SPT) in LaFeAsO (La-1111) and SmFeAsO (Sm-1111) 
single crystals measured by high resolution x-ray diffraction is found to be sharp while the RFeAsO 
(R=La, Nd, Pr, Sm) polycrystalline samples show a broad continuous SPT. 
Comparing the polycrystalline and the single crystal 1111 samples, the critical exponents of the 
SPT are found to be the same 
while the correlation length critical exponents are found to be very different. These results imply 
that the lattice fluctuations in 1111 systems change in samples with different surface to volume ratio 
that is assigned to the relieve of the temperature dependent superlattice misfit strain between 
active iron layers and the spacer layers in 1111 systems. 
This phenomenon that is missing in the AFe$_2$As$_2$ (A=Ca, Sr, Ba) "122" systems, with the same 
electronic structure but different for the thickness and the elastic constant of the spacer 
layers, is related with the different maximum superconducting transition temperature in the 1111 
(55 K) versus 122 (35 K) systems and implies the surface reconstruction in 1111 single crystals.
\end{abstract}

\maketitle
\section{INTRODUCTION}
The natural lattice misfit between first 2D atomic monolayers and second intercalated 
spacer layers forming a 3D superlattice, like in intercalated graphite \cite{bak}, called 
the {\it superlattice misfit strain} (SMS) is known to be a key physical variable to describe 
the physics of these heterostructures at atomic limit. 
The SMS is of wide use in the study of multilayer semiconductor 
heterostructures \cite{Jain}, and of a variety of 3D (2D) bulk systems containing 2D (1D) 
interfaces \cite{forg}. For a given SMS the response of the system depends on the 
difference between the elastic constant of the first and the second layers, their respective 
temperature dependence, and the thickness of spacer layers \cite{Nagao}.
All known high temperature superconductors (HTS), cuprates, diborides and pnictides, are heterostructures 
at atomic limit \cite{bianc1994} made of first atomic superconducting monolayers intercalated
by second layers with variable thickness playing the role of spacers \cite{bianc1993,Rsust}. The SMS is  
a key physical variable controlling the superconducting critical 
temperature, T$_c$, at constant doping in cuprates \cite{bianc2000,pd}, diborides \cite{agre2001}, and 
pnictides \cite{ricci2009}. 
Recently the complex heterogeneity in high Tc superconducting cuprates \cite{Dagotto}, has been related 
to the SMS 
that plays a key role in these functional complex systems \cite{fratini10}.
In pnictides \cite{Zhao,dg,poccia,egami} the T$_c$ at 
constant doping shows very large variation as a function of the SMS that induces the 
deformation of the FeAs lattice, usually measured by the variation of the distance of As 
ion from the Fe plane \cite{Mizu,Zhao_NM}. This deformation is due to the variable SMS 
induced by the variable spacer 
material, since the FeAs layer remains unchanged. The proximity to structural tetragonal-orthorhombic 
phase transition (SPT) in the undoped pnictides 
has been identified as a key feature for high temperature 
superconductivity (HTS) \cite{Tegel,Ni,Wilson-1,Wilson-2,Nomura,Fratini08,Margadonna,Luo,Jesche}. 
The SPT precedes magnetic ordering 
in the parent RFeAsO (1111) compounds \cite{Zhao_NM} 
whereas both transitions occur simultaneously in the AFe$_2$As$_2$ (122) 
compounds \cite{Tegel,Ni,Wilson-1}. 
For the investigation of lattice effects in HTS, it is of high interest to understand the variation 
of the lattice response as function of the elastic constant and thickness of the spacer layers 
in the proximity of the SPT \cite{rt}. 
The SMS is expected to induce a microstrain in the active layers that develops a complex 
lattice structure \cite{bak,forg,Jain,Nagao}. 
The initial studies on the 122 systems indicated the dynamic crystal symmetry breaking to be 
a second order phenomena \cite{Tegel}, however, later studies tend to support a picture of 
a weakly first order transition \cite{Wilson-1} and this topic is an object of active 
investigation \cite{luz, ka}. Here, using high quality single crystals 
together with corresponding polycrystalline powder samples, 
we have measured the SPT in the 1111 systems using high resolution synchrotron x-ray 
diffraction study of the diffraction 
intensities and the line-shape broadening.

\section{EXPERIMENTAL METHODS}
\begin{figure}
\input{epsf}
\epsfxsize 6.25cm
\centerline{\epsfbox{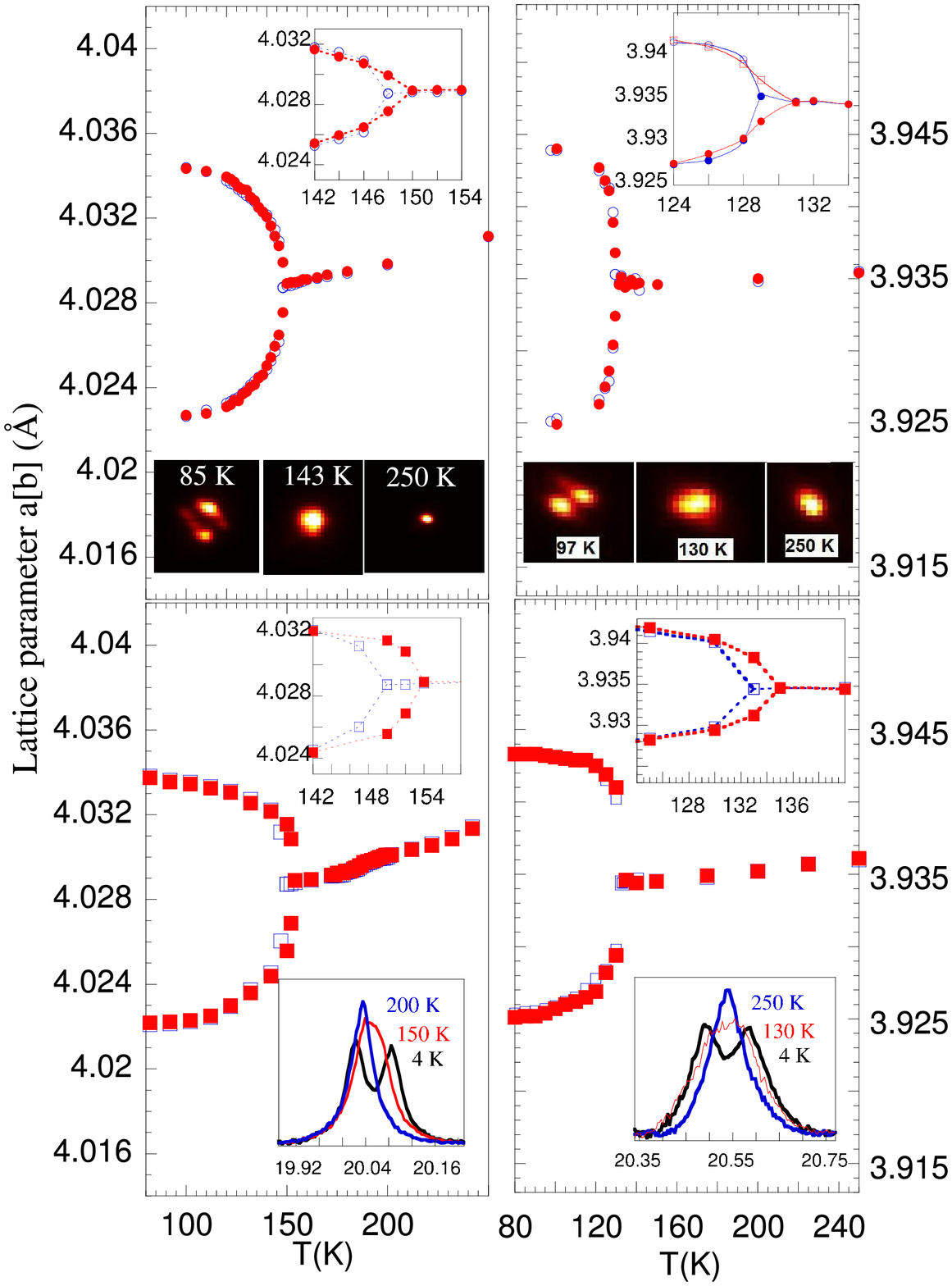}}
\epsfxsize 6.0cm
\centerline{\epsfbox{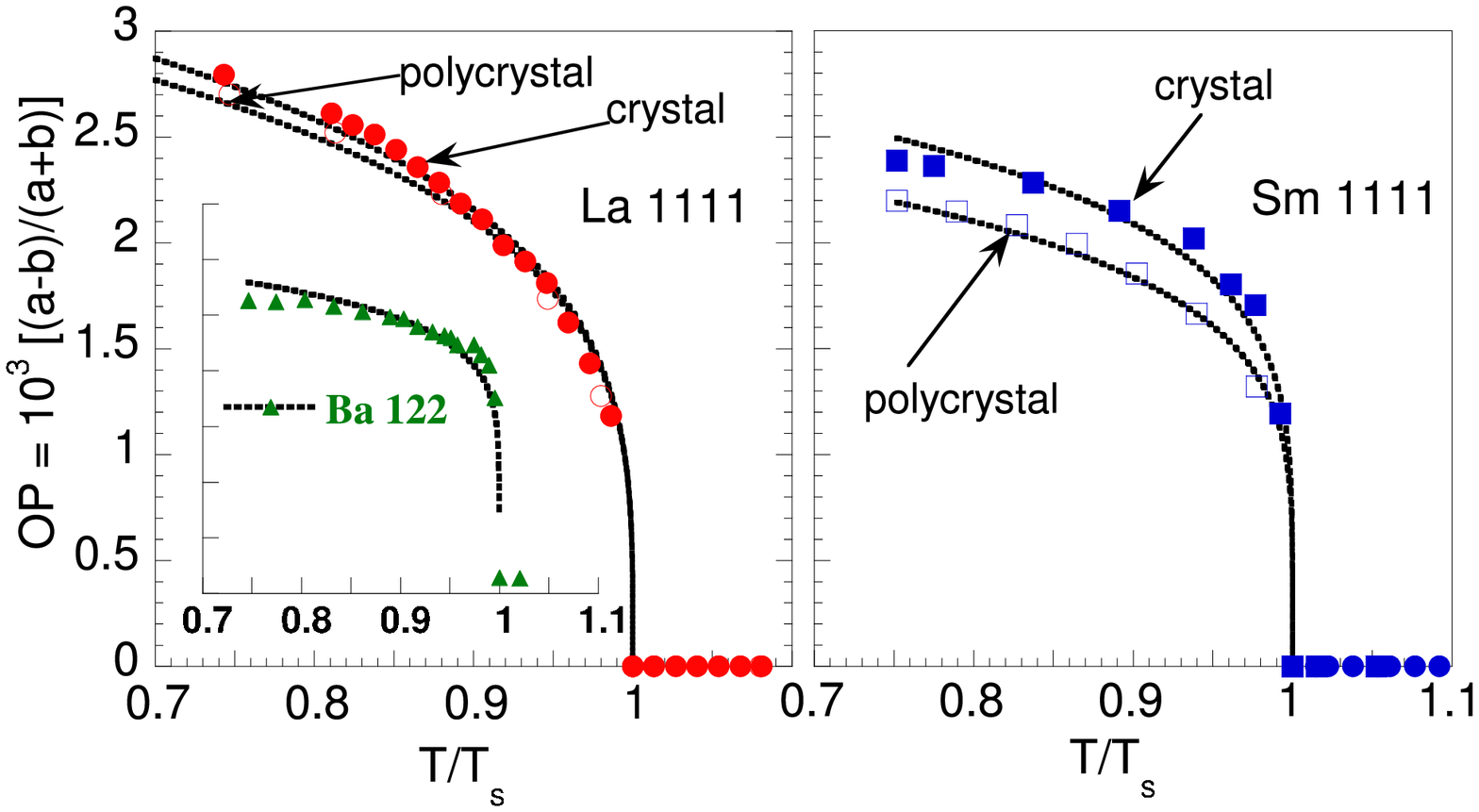}}
\caption{\label{fig:1} (color online) Upper panels show a (b) lattice constant 
of the LaFeAsO (left) and SmFeAsO (right) single crystal samples as a function of temperature 
during cooling (empty blue circles) and warming (filled red circles) cycle. 
Middle panels show the same for the corresponding polycrystalline powder 
samples. Upper insets in these panels show the zoomed region over the SPT indicating 
the presence of a hysteresis, whereas the lower insets in these panels show the 
evolution of the 220 spot/peak during cooling. Lower panel presents the order 
parameters for the single crystals and polycrystalline polycrystalline powders 
during cooling. The order parameter of the BaFe$_2$As$_2$ system taken from 
Ref. \cite{Wilson-1}
is shown in the inset for comparison. }
\end{figure}
The single crystals of the 1111 systems are more unstable and difficult to synthesize 
compared to the 122 compounds. A general method adopted for the synthesis of the 1111 
single crystals is using the cubic anvil high-pressure technique \cite{Karpinski,Zhigadlo}.
Single crystals of SmFeAsO used in this study 
were grown under high pressure in NaCl flux \cite{Karpinski,Zhigadlo} while LaFeAsO single 
crystals were grown under ambient pressure in NaAs flux \cite{Yan}. 
We have used one of the best available Sm-1111 single crystals which have around 6
0 $\mu$m by 60 $\mu$m surface area with $\sim$ 10 $\mu$m thickness. Compared 
to this the La-1111 single crystal was larger, with around 2 mm by 
2 mm surface area and $\sim$ 10 $\mu$m thickness.
The RFeAsO (R=La, Nd, Pr, Sm) polycrystalline samples were 
prepared by high-pressure synthesis method \cite{Zhao}. The x-ray diffraction (XRD) 
data on the single crystal samples were obtained at ELETTRA 
synchrotron radiation facility, Trieste. The data were collected in the K geometry, with a photon 
energy of 12.4 keV using a 2D CCD x-ray detector. 
The sample temperature was varied between 4 and 300 K and stabilized at the set 
point waiting for a temperature gradient
in the sample to be less than 0.1 K. 
All the images measured by single crystal diffraction were properly 
processed using FIT2D program. The XRD measurements on the 
polycrystalline powder samples were performed at the Swiss light source facility at PSI, Zurich. 
The energy resolution was 0.014\%  with photon wavelength 
$\lambda$ =0.495926$\AA$. Data analysis 
were performed with the GSAS suite of
Rietveld analysis programs.  

\section{RESULTS AND DISCUSSIONS}

\begin{figure}
\input{epsf}
\epsfxsize 6.25cm
\centerline{\epsfbox{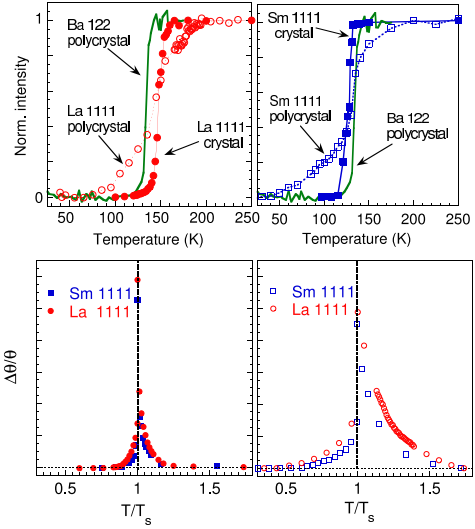}}
\caption{\label{fig:2} (color online) Upper panel: Intensity variation of the 220 peak as a 
function of temperature for the single crystal and polycrystalline powder 
samples of LaFeAsO (marked as La 1111) and SmFeAsO (marked as Sm 1111) systems together 
with the intensity variation of a similar peak observed in the BaFe$_2$As$_2$ system 
(marked as Ba 122 polycrystalline powder). Lower panel: $\Delta\theta/\theta$ of the 220 
peak in the tetragonal 
phase and 400 peak in the orthorhombic phase as a function of T/T$_s$ for the Sm-1111 
and La-1111 single crystals (left panel) and polycrystalline powders 
(right panel). Size of dots shows the dimension of error bar.
}
\end{figure}
Figure 1 shows the temperature dependent variation of the unit cell constants, a and b, 
during the cooling and warming cycles for 
the La-1111 and Sm-1111 single crystals and polycrystalline powders respectively. 
The high resolution x-ray diffraction profile of the 220 reflection of the high 
temperature tetragonal structure (P4/nmm space group) and the 040 and 400 lines of 
the low temperature orthorhombic phase (Cmma space group) of the investigated 
pnictides are shown in Fig. 1. 
To make quantitative analysis, involving the relative intensities and FWHM, 
the peaks were deconvoluted with Gaussian functions.
As one lower the temperature, the diffraction profiles get broader and finally split 
into two distinct peaks clearly 
indicating the SPT (Fig. 1). The nature of the SPT in the 
single crystals and corresponding polycrystalline powders is described by the 
order parameter OP=[(a-b)/(a+b)]$\times$10$^3$, where a and b are the 
lattice constants. In Fig. 1 lower panel, we compare the OP of the 
single crystal samples with the polycrystalline powders. Furthermore, 
the upper insets in all the upper and middle panels of Fig. 1, clearly 
indicate the presence of a hysteresis of the structural phase transition 
in  the 1111 systems.

The OP of the single crystals are sharper than the corresponding polycrystalline 
powders in its approach towards the SPT critical temperature, T$_s$. The data corresponding 
to both single crystal and polycrystalline powder are found to follow a power law 
with the same critical exponent, $\beta$ and values 0.25$\pm$0.02 for La-1111 and 
0.19$\pm$ 0.02 for Sm-1111 respectively. In comparison, the onset of the orthorhombic order 
is reported to have the $\beta$ values 0.103 $\pm$ 0.018 and 0.112$\pm$0.01 in 
BaFe$_2$As$_2$ and EuFe$_2$As$_2$ 
respectively \cite{Tegel,Ni,Wilson-1}. In fact the same analysis, 
taking the data from the literature, for the BaFe$_2$As$_2$  yields
a value 0.136 $\pm$ 0.02 (Fig. 1, inset in the lower panel). 
The difference between the critical exponents of La-1111 and Sm-1111 from 
the (Ba,Eu)Fe$_2$As$_2$ is an index of a different structural coupling of the 
electronic and lattice strain degrees of freedom in the 1111 and 122 
families \cite{Krellner}. The critical exponent of the 
La-1111 system is $\beta$=0.25, 
which is quite different from the mean field calculation of the critical 
exponent $\beta$ =0.5. 
The $\beta$ =0.194 found in Sm-1111 is still lower than 0.25 found in La-1111. 

\begin{figure}
\input{epsf}
\epsfxsize 6.25cm
\centerline{\epsfbox{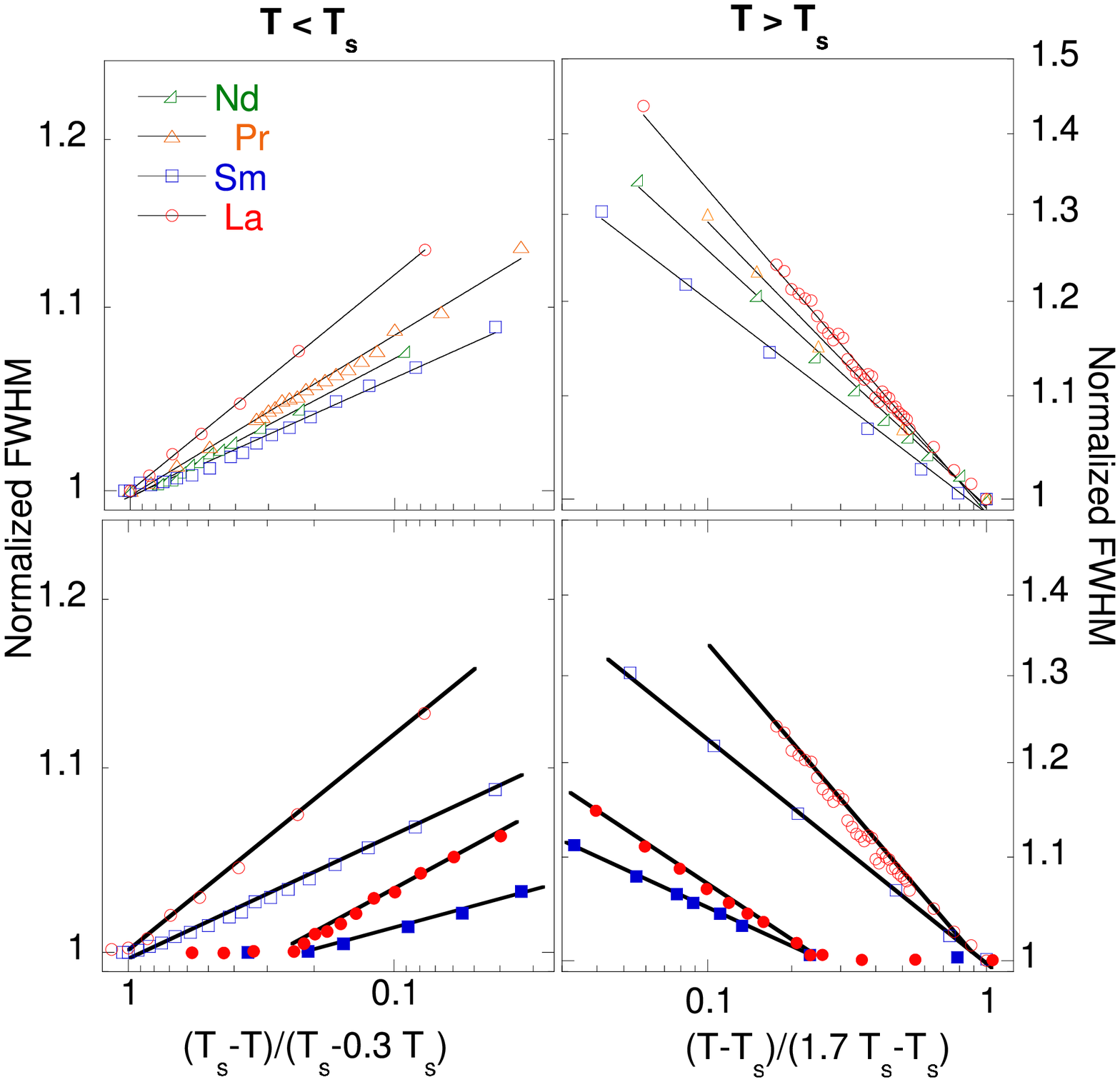}}
\epsfxsize 6.5cm
\centerline{\epsfbox{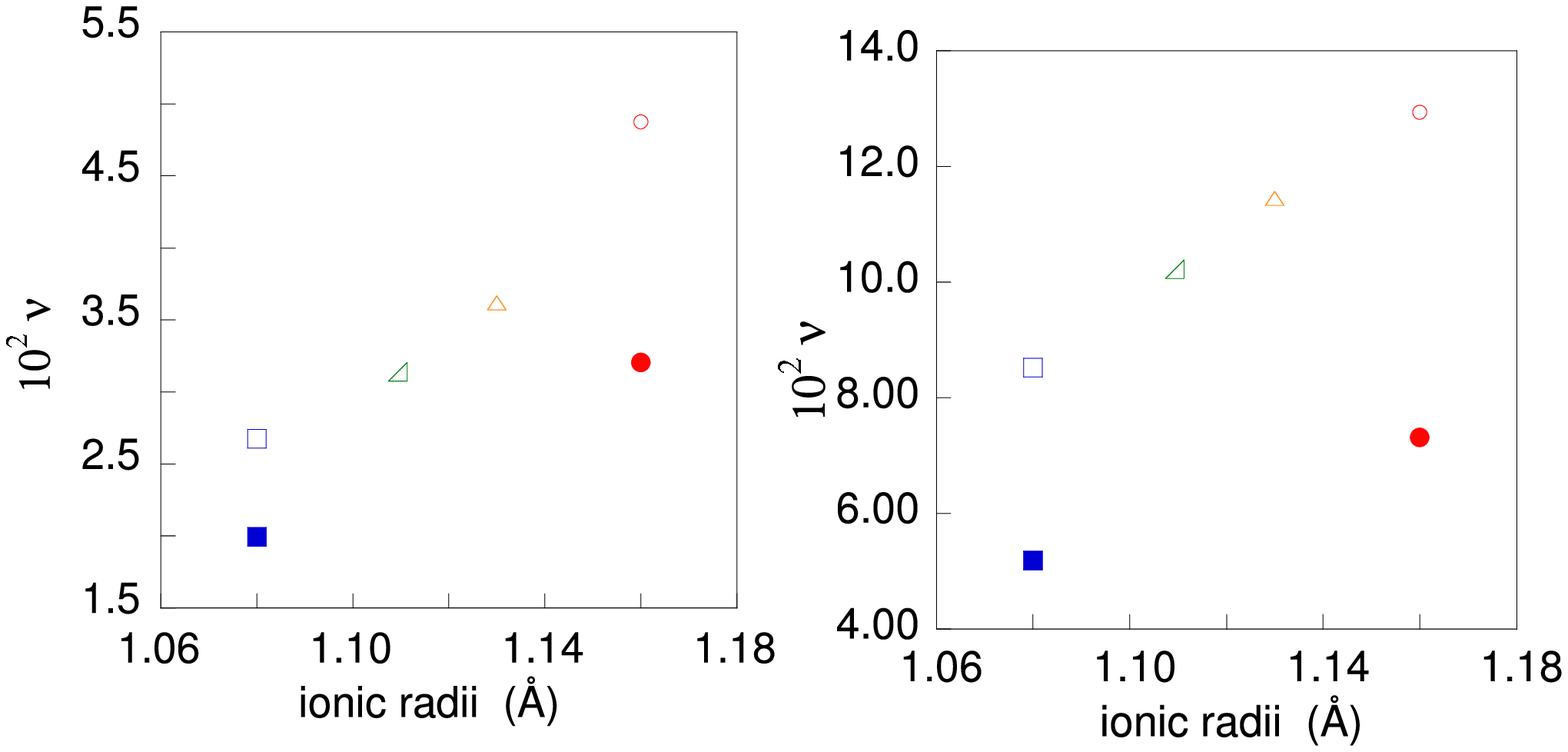}}
\caption{\label{fig:3} (color online) Normalized FWHM of the tetragonal 220 peak and 
corresponding orthorhombic 400 (or 040) peak as a function of temperature for the RFeAsO 
(R=La, Pr, Sm and Nd) systems. T$<$T$_s$ and T$>$T$_s$ are shown in the left and right panels 
respectively. Upper panel compares the behaviour of the polycrystalline 
powder samples. Middle panels compares the normalized FWHM of the single-crystal and 
polycrystalline polycrystalline powder systems. Fit to the data are included as lines. 
The exponent obtained from the fits are compared in the lower panels as a function of 
the rare-earth ionic size. Size of dots shows the dimension of error bar. }
\end{figure}

A comparison of the variation in the intensity of the 220 peak before and after the SPT, 
for the single-crystals and polycrystalline powders show remarkable 
differences, shown in the upper panel of Fig. 2. As evident from the intensity variation, 
the SPT occurs over an extended temperature range of about 90 K for the 
polycrystalline powders, whereas the SPT process is confined within a window 
of around 20 K in single crystals. Upon cooling-warming cycles, a similar 
temperature hysteresis, as seen in the lattice constant, is also seen in the 
intensity plots (not shown). From Fig. 2, it is clear that the SPT 
behaviour in 1111 single crystals and 122 systems are quite similar. As one approaches 
the structural transition temperature, the polycrystalline sample due to its 
finite size increasingly becomes vulnerable to the lattice fluctuations 
leading to an overall broadening of the transition region, resulting in an 
effective increase of the T$_s$ values of the polycrystalline powders compared to 
single crystals (see Fig. 2). The fact that this effect is seen only in the 1111 
systems, and not in the 122, implies that the origin of this effect is due to the 
presence of the spacer layer in the former. The large difference of the lattice fluctuations
 near a structural phase transition of 1111 samples with different surface to volume ratio 
show a lattice instability much bigger compared to the 122 systems. It is instructive to 
compare the evolution of the full width at half maximum (FWHM) of the tetragonal 200 peak 
and the corresponding orthorhombic peaks (400 or 040), which is shown in Fig. 2 
lower panel. Approaching the T$_s$, the FWHM has longer tail for polycrystalline 
powder than in single crystal samples. It is in fact well known that the widths 
of diffraction lines are inverse to the sizes of crystallites formed during 
the material synthesis, and that these lines are broadened by 
microstrain \cite{Nagao,Frey}. 
The difference in the SPT behavior 
in the single crystal and polycrystalline powder can be understood invoking the 
idea of larger crumbling of the micro-crystallites of the polycrystalline 1111 samples 
in comparison to the single
crystals as one approaches the SPT temperature.

In Fig. 3, we plot the normalized FWHM with the normalized temperature for the polycrystalline 
powders and single crystals. 
The results for the PrFeAsO and NdFeAsO polycrystalline powder samples are also 
shown. For temperature below T$_s$, the normalization is done by taking the value 
of the FWHM at 0.3$\times$T$_s$ to unity (Fig. 3 lower panel), while for temperatures 
above T$_s$, the normalization is done by taking the FWHM values at 1.7$\times$ T$_s$ to unity. 
The correlation length $\xi$ 
(has an inverse relation with FWHM, see Ref. \cite{Fisher,patt}) of the line-shape approaching 
T$_s$ is well described by a power law $\xi^{-1}= t^{v}$ where t is the reduced temperature 
(defined in Fig. 3). Although both single crystals and polycrystalline powders 
are found to follow the $\xi^{-1}= t^{v}$ power law, the corresponding exponent, $\nu$, 
for the polycrystalline powder and the single crystal are found to be very different 
for the identical system, the later being 4 times higher. Such powder 
law fits for the polycrystalline powders of La-1111, Pr-1111, Sm-1111 and Nd-1111 are 
shown in the upper panels of Fig. 3. 
In the case of the polycrystalline powder samples, 
the exponent increases almost linearly with increasing rare-earth ionic size 
(see Fig. 3 lower panels). 
Marked difference in the correlation length exponents 
observed in the case of the 
polycrystalline powder and the corresponding single 
crystals is an evidence of the crystallite size dependent SMS effects in
the 1111 system. However both 1111 materials 
(grown with two different procedures) show different lattice fluctuations going from
microcrystallines of powders (diameter less than 1 micron) to larger single crystals. 
The difference between the small grains and large single crystals is attributed to the 
difference between the elastic constant of active FeAs and RO spacer layers.
In 1111 policrystalline systems the surface of grains is 
expected to be different since the surface layer has a different elastic strain 
compared to the layers in the bulk.
On the contrary, the 122 systems show similar lattice response for the small (policrystalline powder) 
and for large crystals indicating that the surface to volume ratio does not play a significant 
role in 122 systems and the elastic stress due to the natural interlayer misfit is different. 
This difference in the 
lattice response could be related to the unexplained difference of the superconducting critical 
temperature between 1111 and 122 samples having similar electronic structure.

\section{CONCLUSIONS}
In conclusion, the structural phase transition in the La-1111 and Sm-1111 appears 
to be a case of intermixing of first and second order 
transition that in correlated materials is not rare. The comparison between the  
x-ray diffraction data for the polycrystalline 1111 samples and the single crystals shows a 
relevant differences as one approach the SPT temperature. 
This is assigned to an elastic response dependence of the surface to volume ratio of the sample.
Difference in the $\beta$ exponent and the temperature dependence of the single crystal 
and polycrystalline powder data underline the importance of the superlattice misfit 
strain  \cite{bianc2000,fratini10} in the phase diagram and for the functional 
properties \cite{Dagotto} of these heterostructures at atomic limit.
The 122 systems on the contrary show the same lattice fluctuations in micro-crystals and 
large-crystals. This difference (between the 1111 and 122) is assigned to the difference 
between the elastic constant of the spacer layers in the two systems.
The electronic structure of 1111 and 122 systems is very similar, so this difference in the 
dynamical response between the 1111 and 122 systems may explain the increase of the T$_c$, 
from 35 K in 122 to 55 K in 1111 systems in fact the misfit strain has been 
proposed to be the key term determining the critical multiscale phase separation in doped 
high temperature superconductors giving the so called superstripes scenario.\cite{fratini10,stripes} 

\section*{ACKNOWLEDGMENTS}
We thank Z.-A. Ren and Z.-X. Zhao for providing the powder samples, the X04SA beamline staff of 
Swiss light source, Zurich, for help, the Paul Scherrer Institute, Zurich, for support
of European researchers and N.L. Saini for 
discussions. This project is supported by the Sapienza University 
of Rome. The work at the ETH Zurich was supported by the Swiss National 
Science Foundation through the NCCR pool MaNEP. Ames laboratory is operated 
for US Department of Energy by Iowa State University under contract 
number DE-AC02-07CH11358.

\end{document}